\documentclass[twocolumn]{aastex6}

\newcommand{\oiii}{[O{\,\sc iii}]\,\,}

\begin{document}


\title{No sign of strong molecular gas outflow in an infrared-bright dust-obscured galaxy with strong  ionized-gas outflow}


\author{Yoshiki Toba 		\altaffilmark{1,2},
		Shinya 	Komugi 		\altaffilmark{3},
		Tohru Nagao 		\altaffilmark{2},
		Takuji Yamashita 	\altaffilmark{2},
		Wei-Hao Wang		\altaffilmark{1},
		Masatoshi Imanishi	\altaffilmark{4,5},
		Ai-Lei Sun			\altaffilmark{6,1}
		}	
\affil{}  			  
  \altaffiltext{1}{Academia Sinica Institute of Astronomy and Astrophysics, PO Box 23-141, Taipei 10617, Taiwan}
  \email{toba@asiaa.sinica.edu.tw}
  \altaffiltext{2}{Research Center for Space and Cosmic Evolution, Ehime University, 2-5 Bunkyo-cho, Matsuyama, Ehime 790-8577, Japan} 
  \altaffiltext{3}{Kogakuin University, 2665-1 Nakanocho, Hachioji, Tokyo 192-0015, Japan}
  \altaffiltext{4}{National Astronomical Observatory of Japan, 2-21-1 Osawa, Mitaka, Tokyo 181-8588, Japan}
  \altaffiltext{5}{Department of Astronomical Science, The Graduate University for Advanced Studies (SOKENDAI), Mitaka, Tokyo 181-8588, Japan}
  \altaffiltext{6}{Department of Physics and Astronomy, Bloomberg Center, Johns Hopkins University, Baltimore, MD 21218, USA}



\begin{abstract}
We report the discovery of an infrared (IR)-bright dust-obscured galaxy (DOG) that shows a strong ionized-gas outflow but no significant molecular gas outflow.
Based on detail analysis of their optical spectra, we found some peculiar IR-bright DOGs that show strong ionized-gas outflow ([O{\,\sc iii}]$\lambda$5007) from the central active galactic nucleus (AGN).
For one of these DOGs (WISE J102905.90+050132.4) at $z_{\rm spec} = 0.493$, we performed follow-up observations using ALMA to investigate their CO molecular gas properties.
As a result, we successfully detected $^{12}$CO($J$=2--1) and $^{12}$CO($J$=4--3) lines, and the continuum of this DOG.
The intensity-weighted velocity map of both lines shows a gradient, and the line profile of those CO lines is well-fitted by a single narrow Gaussian, meaning that this DOG has no sign of strong molecular gas outflow.
The IR luminosity of this object is $\log\,(L_{\rm IR}/L_{\sun})$ = 12.40  that is classified as ultraluminous IR galaxy (ULIRG).
We found that (i) the stellar mass and star-formation rate relation and (ii) the CO luminosity and far-IR luminosity relation are consistent with those of typical ULIRGs at similar redshifts.
These results indicate that the molecular gas properties of this DOG are normal despite that its optical spectrum showing a powerful AGN outflow.
We conclude that a powerful ionized-gas outflow caused by the AGN does not necessarily affect the cold interstellar medium in the host galaxy at least for this DOG.
\end{abstract}

\keywords{galaxies: active  --- infrared: galaxies  --- radio lines: galaxies  --- radio continuum: galaxies}



\section{Introduction} 
\label{Intro}
It has been revealed that almost all galaxies have a supermassive black holes (SMBHs) with mass of $10^{5-10} M_{\odot}$ at their nuclei.
The masses of SMBHs are tightly correlated with those of the spheroid components of their host galaxies,  indicating that SMBHs and their host galaxies coevolve \citep[e.g.,][]{Magorrian,Marconi,Kormendy,McConnell,Sun_13,Woo_13,van}.
How did galaxies and their SMBHs coevolve throughout the history of the Universe?
It is widely believed that a key process to solve this puzzle is so-called active galactic nucleus (AGN) feedback; radiation, winds, and jets from the AGN can interact with its interstellar medium (ISM), and this can lead to ejection or heating of the gas. 
Observational and theoretical investigations have indicated that these powerful outflows resulting from AGN feedback regulate star formation (SF) and even AGN activity, and could control the co-evolution of galaxies and SMBHs \citep[e.g.,][]{Silk,Di,Cano-Diaz,Fabian,King,Bieri}.

In this work, we focus on infrared (IR)-bright dust-obscured galaxies \citep[DOGs:][]{Dey,Toba_15} characterized by an extreme optical/IR color, as a laboratory to probe the AGN feedback phenomenon.
This is because IR-bright DOGs are expected to be at the maximum phase of AGN and SF activities in the framework of the galaxy--SMBH co-evolution behind a large amount of dust \citep[e.g.,][]{Hopkins_08,Narayanan,Ricci,Toba_17d}.
We have performed a systematic IR-bright DOGs search and investigated their statistical properties \citep{Toba_15,Toba_16,Toba_17a,Toba_17b} using the optical and IR data including the Sloan Digital Sky Survey \citep[SDSS:][]{York} and {\it Wide-field Infrared Survey Explorer} ({\it WISE}: \citealt{Wright}).
Recently, \cite{Toba_17c} found several peculiar IR-bright DOGs that show a strong  ionized-gas outflow in the SDSS spectrum.
The velocity offset with respect to the systemic velocity measured from the stellar absorption lines and the velocity dispersion of the [O{\,\sc iii}]$\lambda$5007 line are 500 -- 2000 km s$^{-1}$.
Since most of them are AGN dominated DOGs with IR luminosity exceeding $10^{12}$ $L_{\sun}$ (that are termed ultraluminous IR galaxies, ULIRGs: \citealt{Sanders_96}), given the high luminosity of the AGN, it is likely that these ionized outflows are powered by AGN radiation or wind.
However, although the AGN-driven outflow of ionized-gas has been confirmed, whether or not there is molecular gas associated with outflows is still unclear.
Observations of molecular gas are useful not only to check the presence of outflowing molecular gas but also to investigate its kinematics and energetics \citep[e.g.,][]{Feruglio_10,Alatalo,Sturm,Cicone_14,Garcia-Burillo,Sakamoto,Sun,Feruglio_15,Gallimore,Querejeta,Imanishi}.
In particular, because stars form in dense molecular clouds, AGN feedback could only regulate SF if it has impact on the molecular gas content. 
The Atacamma Large Millimeter/submillimeter Array (ALMA) provides an avenue to realize the above requirements, as well as assessing the content of the molecular gas reservoir.
Since molecular gas is the ingredient of SF, our observations with ALMA are important to understand the connection between AGN feedback on the SF.

In this paper, we present follow-up observations of a DOG showing a strong \oiii gas outflow using ALMA with band 4 and 7 to probe its $^{12}$CO($J$=2--1), $^{12}$CO($J$=4--3), and continuum emissions.
Throughout this paper, the adopted cosmology is a flat Universe with $H_0$ = 70 km s$^{-1}$ Mpc$^{-1}$, $\Omega_M$ = 0.3, and $\Omega_{\Lambda}$ = 0.7. Unless otherwise noted, all magnitudes refer on the AB system and a \cite{Chabrier} initial mass function (IMF) is assumed.

\section{Data and analysis} 
\label{DA}

\subsection{Target selection}
The target, WISE J102905.90+050132.4 (hereafter WISE1029), for the ALMA observations was selected from the DOG sample in \cite{Toba_16}.
They first selected 67 IR-bright DOGs with $(i - [22])_{\rm AB} > 7.0$ and flux density at 22 $\micron$ $>$ 3.8 mJy from the SDSS spectroscopic catalog \citep{Alam} and ALLWISE catalog \citep{Cutri}.
\cite{Toba_17c} then narrowed down to 36 objects with 0.05 $< z <$ 1.02 that clearly have \oiii in their SDSS spectra and performed a detail analysis for [O{\,\sc iii}].
In particular, they measured the velocity offset ($v_{\rm [OIII]}$) with respect to the systemic velocity  and the velocity dispersion ($\sigma_{\rm [OIII]}$) as indicators of outflowing gas \citep[see ][]{Bae,Bae_16,Woo,Bae_17,Woo_17,Toba_17c}.
As a result, WISE1029 at $z_{\rm spec}$ = 0.4930 was discovered.
Its $v_{\rm [OIII]}$ and $\sigma_{\rm [OIII]}$ are $-1485.0 \pm 207.7$ km s$^{-1}$ and $986.9 \pm 146.6$ km s$^{-1}$, respectively, indicating that  ionized-gas is outflowing from the WISE1029 (see object with ID = 14 in Table 1, 2, and Figure 14 of \citealt{Toba_17c}, for more details).
Since WISE1029 is an extreme case in terms of $v_{\rm [OIII]}$ and $\sigma_{\rm [OIII]}$ among our DOG sample and it is observable from ALMA, we select this object as a target of ALMA observation.
The observed properties of WISE1029 are summarized in Table \ref{Table}.

\begin{center}
\begin{table}[h]
\begin{footnotesize}
\renewcommand{\thetable}{\arabic{table}}
\caption{Observed properties of WISE1029.}
\label{Table}
\begin{tabular}{lr}
\tablewidth{0pt}
\hline
\hline
WISE J102905.90+050132.4		&								\\
\hline
R.A. (SDSS) [J2000.0] 			& 	10:29:05.90					\\
Decl. (SDSS) [J2000.0]			& 	+05:01:32.42 				\\
Redshift \citep{Toba_17c}		&	0.4930						\\
Band 4 continuum (146.6 GHz) [mJy]					&	0.18 $\pm$ 0.02 \\
Band 7 continuum (303.9 GHz) [mJy]					&	0.85 $\pm$ 0.09 \\
$S_{\rm CO (2-1)}$ $\Delta v$ [Jy km s$^{-1}$]		&	1.90 $\pm$ 0.13 \\
$S_{\rm CO (4-3)}$ $\Delta v$ [Jy km s$^{-1}$]		&	4.66 $\pm$ 0.47 \\
FWHM$_{\rm CO (2-1)}$ [km s$^{-1}$]					&	336.3 $\pm$ 32.4 \\
FWHM$_{\rm CO (4-3)}$ [km s$^{-1}$]					&	373.1 $\pm$ 15.1 \\
$\log\,L_{\rm FIR}$ (40-120 $\micron$) [$L_{\sun}$]			&	$11.83^{+1.63}_{-0.36}$	\\
$\log\,L_{\rm IR}$ (8-1000 $\micron$) [$L_{\sun}$]			&	$12.40^{+0.70}_{-0.17}$	\\
$\log\,L^{\prime}_{\rm CO (2-1)}$ [K km s$^{-1}$ pc$^2$]	&	$10.25 \pm  0.03$	\\
$\log\,L^{\prime}_{\rm CO (4-3)}$ [K km s$^{-1}$ pc$^2$]	&	$10.04 \pm  0.04$   \\
$r_{\rm 42}$											&	0.61 $\pm$ 0.07	\\
$\log$ $M_*$ [$M_{\sun}$]								&	$10.8^{+0.02}_{-0.06}$\\
$\log$ SFR [$M_{\sun}$ yr$^{-1}$]						&	$2.11^{+0.64}_{-0.28}$\\
$\log$ $M_{\rm dust}$ [$M_{\sun}$]						&	8.5\\
$\log$ $M_{\rm gas}$ [$M_{\sun}$]						&	10.2 \\
$M_{\rm gas}$ 	/ 	 $M_{\rm dust}$						&	53\\
\hline
\end{tabular}
\end{footnotesize}
\end{table}
\end{center}

\begin{figure*}
\centering
\includegraphics[width=0.95\textwidth]{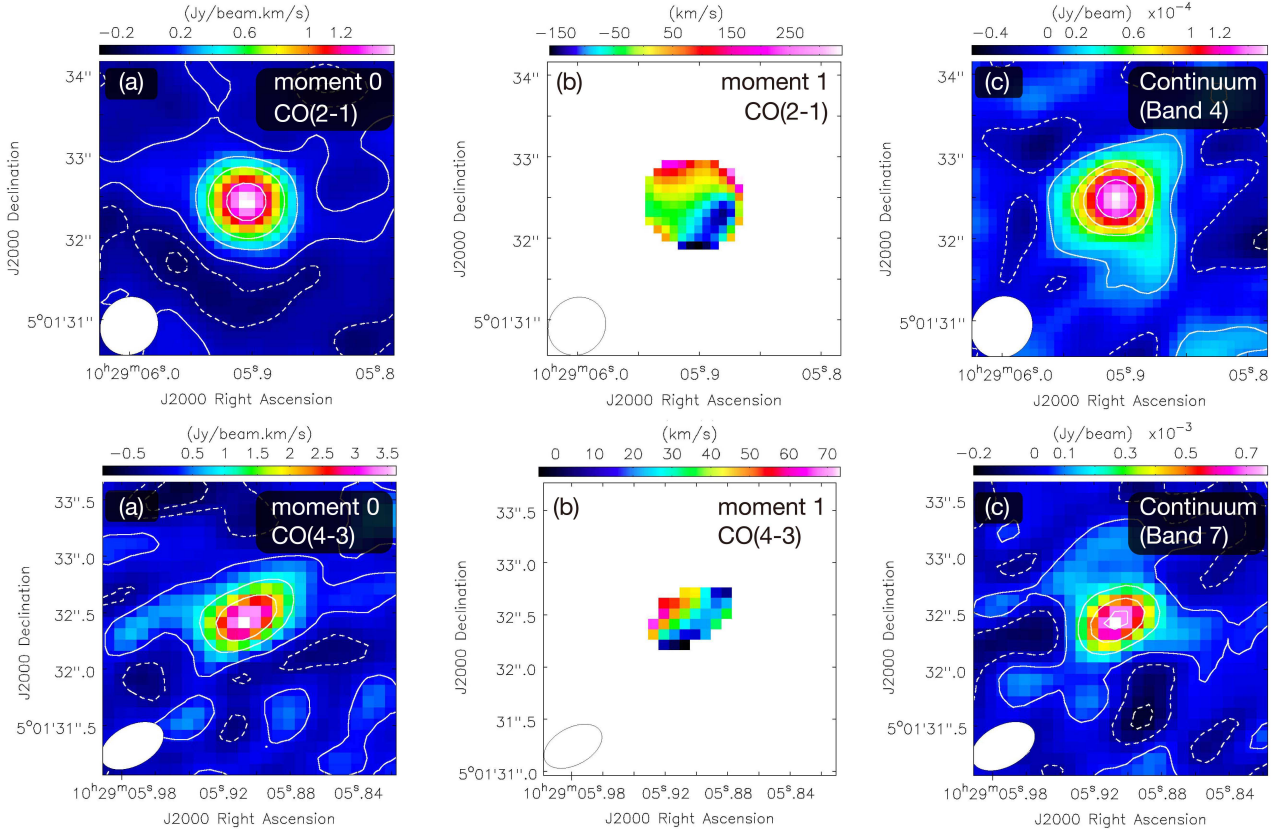}
\caption{(a) Integrated intensity (moment 0) maps of CO lines. The velocity range used to create the image is  3640 and 1809 km s$^{-1}$ centered at CO(2--1) and CO(4--3) line, respectively. The 1$\sigma$ noise level of the image is 0.06 and 0.21 Jy beam$^{-1}$ km s$^{-1}$ for CO(2--1) and CO(4--3), respectively. The contour levels are -2, -1, 1, 5, 10, and 20$\sigma$ for CO(2--1) image and -2, -1, 1, 5, 10, and 15$\sigma$ for CO(4--3) image. (b) Intensity-weighted mean velocity (moment 1) map of CO lines where components with $> 5\sigma$ are shown. (c) Intensity map of continuums at 146.6 and 303.9 GHz. Top and bottom panels show CO(2--1) and CO(4--3), respectively. The 1$\sigma$ noise level of the image is $0.12 \times 10^{-4}$ and $0.46 \times 10^{-4}$ Jy beam$^{-1}$ for band 4 and 7, respectively. The contour levels are -3, -1, 1, 5, 7, and 10$\sigma$ for band 4 image and -3, -1, 1, 5, 10, and 15$\sigma$ for band 7 image.}
\label{map}
\end{figure*}

\subsection{ALMA observations}
ALMA observations of the CO($J$=2--1) and CO($J$=4--3) lines were carried out as a part of our ALMA Cycle 3 program (\#2015.1.00199.S PI: Y.Toba).  
The CO(2--1) line was observed in band 4 on 2016 July 14 with the C40-4 configuration, and CO(4--3) in band 7 on 2016 June 22 with the C40-5 configuration.  
For both runs, J1058+0133 was used as the primary flux calibrator and the bandpass calibrator, and J1038+0512 as the phase calibrator.  
Flux uncertainties are consistent with ALMA specifications (5\% in band 4 and 10\% in band 7).  
For both runs, one of the four spectral windows were centered on the molecular line, with 1875 MHz bandwith (corresponding to 3640 $\mathrm{km\ s^{-1}}$ in band 4 and 1809 $\mathrm{km\ s^{-1}}$ in band 7), and 1953 kHz resolution.  
Three other spectral windows were set to observe the continuum, each with 1875 MHz bandwidth.  
The data were calibrated with the ALMA pipeline r38377 for band 4, and r36660 for band 7.  
Continuum subtraction was done using the velocity offset range of -1776 to -556 km s$^{-1}$ and 609 to 1829.63 km $s^{-1}$ for band 4, and -818 to -488 km s$^{-1}$ and 596 to 878 km s$^{-1}$ for band 7.  Both upper and lower panels (a) in Figure \ref{map} are continuum subtracted.
Images were produced by CLEAN with briggs weighting and robust=2 parameter.  
The resulting images have angular resolution of $0^{\prime \prime}.74 \times 0^{\prime \prime}.68$ with position angle of --42 deg. in band 4, and $0^{\prime \prime}.60 \times 0^{\prime \prime}.36$ with position angle of --62 deg. in band 7.

\section{Results} 
\label{RD}

\subsection{The properties of CO molecular gas}
\label{CO}
    
Integrated intensity (moment 0) maps of CO(2--1) and CO(4--3) lines, and continuum images at 146.6 (band 4) and 303.9 GHz (band 7) are shown in Figure \ref{map}.
CO lines are clearly detected with peak signal-to-noise ratio (SN) = 25 for CO(2--1) image and SN = 17 for CO(4--3) image.
The continuums are also detected with peak SN = 13 for band 4 image and SN = 17 for band 7 image.
We measured the flux densities of CO lines and continuums based on 2-D Gaussian fitting without adopting any sigma clipping, and the resultant values are listed in Table \ref{Table}.
Figure \ref{map} also shows the intensity-weighted mean velocity (moment 1) maps of CO lines.
The velocity gradients for both lines can be seen, indicating that CO molecular gas in the DOG is unlikely to be strongly disturbed.

\begin{figure}
\centering
\includegraphics[width=0.45\textwidth]{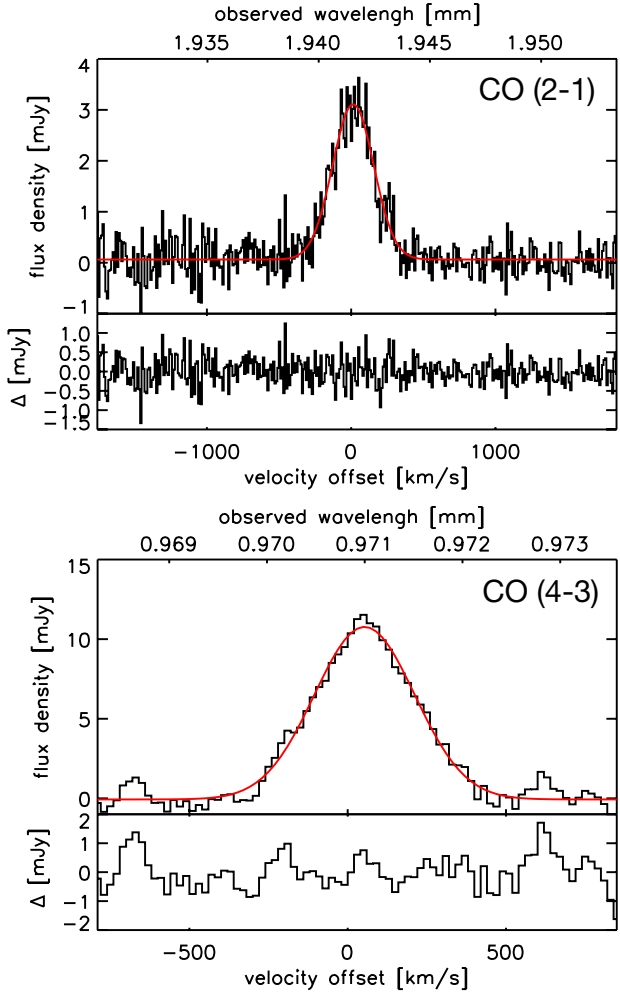}
\caption{CO(2--1) (top) and CO(4--3) (bottom) spectra of WISE1029 obtained with the ALMA. The best fit lines with Gaussian are over-plotted as red solid lines. The velocity is relative to the systemic velocity measured by the fitting with a stellar component in \cite{Toba_17c}. The residual between data and best fit Gaussian is also shown in lower panel of each spectrum.}
\label{line}
\end{figure}

Figure \ref{line} shows the ALMA spectra of the CO(2--1) and CO(4--3) emission lines measured in 1$\arcsec$ diameter apertures.
The velocity is relative to the systemic velocity measured by the fitting with a stellar component in \cite{Toba_17c}. 
We fit each spectrum with a single Gaussian and see if they have broad wing with a full width at half maximum (FWHM) $>$ 500 km s$^{-1}$ that is an indicator of strong outflow \citep[e.g., ][]{Cicone_12,Cicone_14}.
Both CO lines are well-fitted by a single Gaussian with a FWHM of $\sim$350 km s$^{-1}$, which is close to a typical value of local (U)LIRGs \citep[e.g.,][]{Yamashita}.
The residual indicates that there are no broad wings in both CO lines.
The measured parameters of the fitting are tabulated in Table \ref{Table}.
These results indicate that this DOG does not show a significant molecular gas outflow in spite of showing a powerful ionized-gas outflow.
We will discuss an interpretation of this fact in Section \ref{Disc}. 

We note that there still is possibility that CO molecular gas that is not associated with global gas distribution is weakly outflowing.
However, the line flux of the broad component is 1/5 -- 1/10 of the total line flux at the peak for many cases \citep{Alatalo,Cicone_14}. 
Since peak SN of CO lines in this work is about 20, it is hard to detect broad component in our data, even if weak outflowing gas exists (see also Section \ref{Dust}).
In order to discriminate these weak outflowing gas from spectrum, data with high peak SN ($>$ 50) are required.
It is difficult to rule out the existence of a weak CO outflow with the current data, which is the scope of the future work. 

\subsection{The stellar mass and star formation rate relation}
\label{MS}
Next, we investigate the relation between stellar mass ($M_*$) and star formation rate (SFR) of WISE1029 in order to assess if the SF is affected relative to the stellar content by the strong ionized outflow or other AGN feedback mechanisms. 
We perform SED fitting taking advantage of the two new ALMA photometries at 987 and 2045 $\micron$.  
This method using code SEd Analysis using BAyesian Statistics ({\tt SEABASs}\footnote{http://xraygroup.astro.noa.gr/SEABASs/}: \citealt{Rovilos}) is described in details in \cite{Toba_17b,Toba_17d}.
This fitting code provides a best-fit SED as a combination of stellar, AGN, and SF component and IR luminosity of each component.
\begin{figure}
\centering
\includegraphics[width=0.45\textwidth]{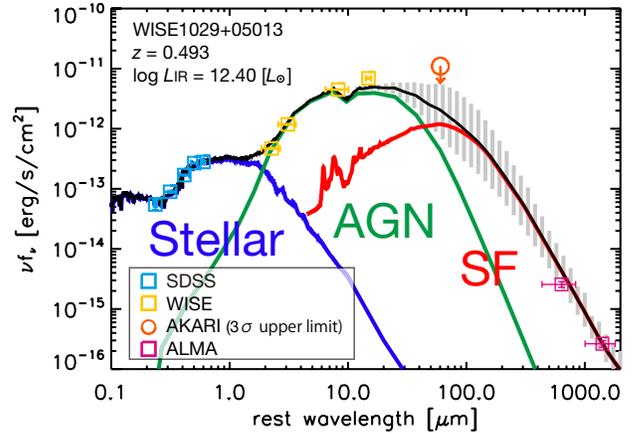}
\caption{SED of WISE1029. The blue, yellow, orange, and pink squares represent the data from SDSS, {\it WISE}, {\it AKARI} (3$\sigma$ upper limit), and ALMA, respectively. The contribution from the stellar, AGN, and SF components to the total SEDs are shown as blue, green, and red lines, respectively. The black solid line represents the resultant SED. The gray shaded region represents the uncertainty of the best fit SED.}
\label{SED}
\end{figure}
We performed the SED fitting using 12 photometric points including 90 $\micron$ data (3$\sigma$ upper limit) obtained from {\it AKARI} far-IR (FIR) all-sky survey \citep{Murakami,Kawada,Yamamura}.
We estimated stellar mass ($M_*$), total IR luminosity ($L_{\rm IR}$ (8-1000 $\micron$)), and IR luminosities contributed from each component.
The SED with a best fit model of WISE1029 is shown in Figure \ref{SED}.
The derived IR luminosity is $\log (L_{\rm IR}/L_{\sun}) = 12.40^{+0.70}_{-0.17}$, which means that WISE1029 is an ULIRG.
The energy contribution of AGN to the total IR luminosity, $L_{\rm IR}$(AGN)/$L_{\rm IR}$ is about 54\%.
Note that the estimated IR luminosity has large uncertainty because we lack the rest-frame FIR data, and thus AGN contribution to the total IR luminosity has also large uncertainty.
We should keep in mind that the current data set still cannot constrain well the FIR SED.
The SFR is converted from $L_{\rm IR}$ (SF) using a relation of $\log \,{\rm SFR} = \log \, L_{\rm IR}$ (SF) - 9.966 \citep{Salim} \citep[see also][]{Toba_17b}.
The measured stellar mass and SFR are listed in Table \ref{Table}.

   \begin{figure}
   \centering
   \includegraphics[width=0.45\textwidth]{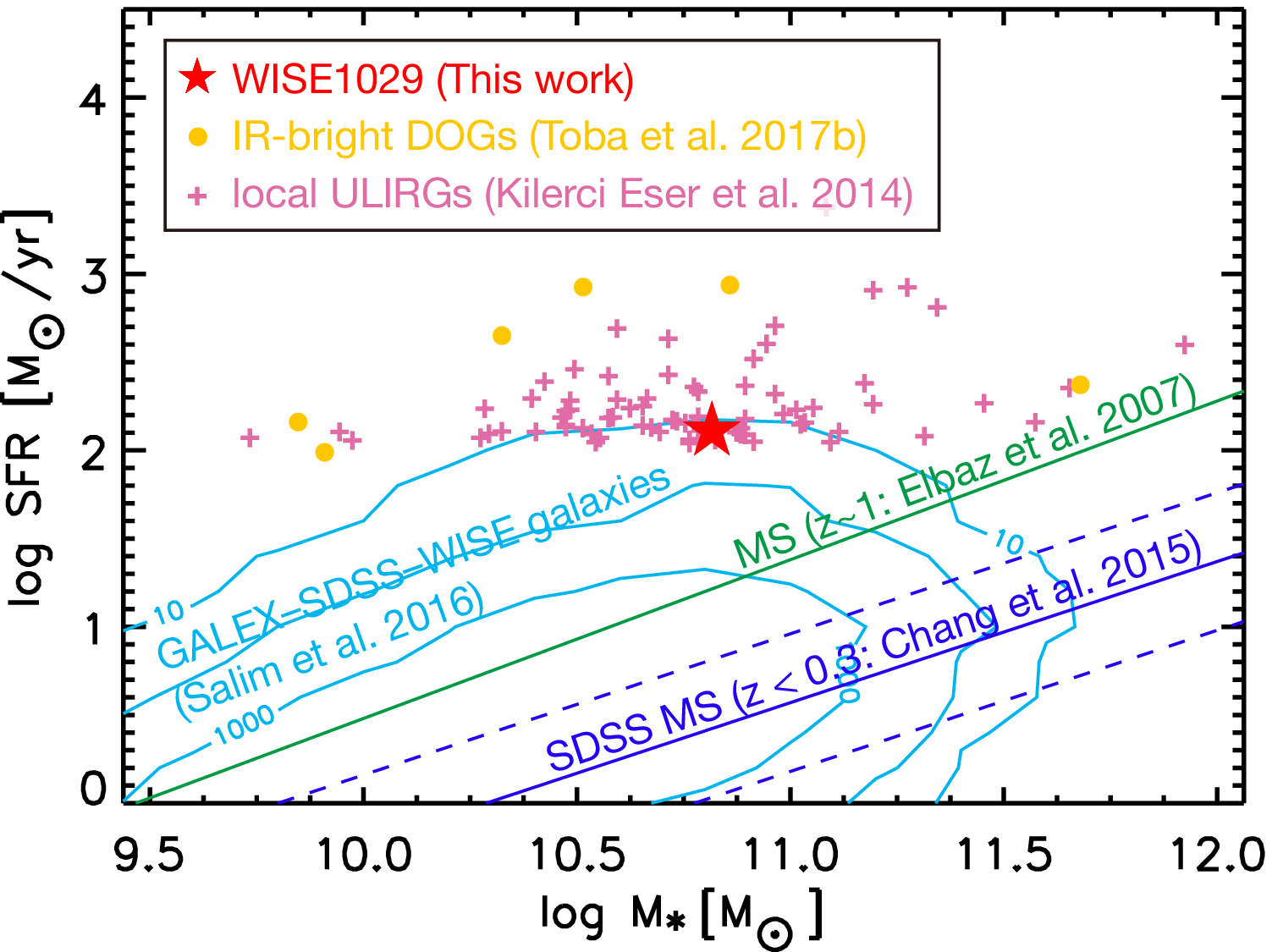}
   \caption{
   Stellar mass and SFR for WISE1029 (red star). The yellow circles represent those for IR-bright DOGs detected with {\it AKARI} or {\it IRAS} \citep{Toba_17b}. The pink crosses represent AKARI-selected ULIRGs at $0 < z < 0.5$ \citep{Kilerci}. The blue solid line is main sequence (MS) of normal SF galaxies selected from the SDSS \citep{Chang} with scatter of 0.39 dex (blue dotted line). The cyan contours represent SFR--$M_*$ relation for a sample of GALEX--SDSS--WISE Legacy Catalog \citep[GSWLC: ][]{Salim} at $z < 0.3$. The bin size is 0.2 $\times$ 0.2 in the units given in the plot. The green line is MS of SF galaxies at $z$ = 1 \citep{Elbaz}.
   }
	\label{MSFR}
   \end{figure}
Figure \ref{MSFR} shows the relation between stellar mass and SFR relation for WISE1029, the main-sequence (MS) sample for star forming galaxies selected by the SDSS and {\it WISE} \citep{Chang}, and those selected by the Galaxy Evolution Explorer ({\it GALEX}) satellite \citep{Martin}, SDSS, and {\it WISE} \citep{Salim}.
The stellar mass and SFR of IR-bright DOGs detected by {\it Infrared Astronomical Satellite} \citep[{\it IRAS}: ][]{Neugebauer,Beichman} and/or {\it AKARI} presented by \cite{Toba_17b} is plotted.
The stellar mass and SFR of MS presented by \cite{Elbaz} for star forming galaxies at $z$ = 1 and those of ULIRGs at $0 < z < 0.5$ presented by \cite{Kilerci} are also shown in Figure \ref{MSFR}.
Note that we corrected a possible offset of stellar masses discussed in Section 4.2.1 of \cite{Toba_17b} for the local SDSS sample \citep{Salim} and ULIRG sample \citep{Kilerci}.
We also corrected the stellar mass and SFR in the literature to those assumed Chabrier IMF if needed \citep[see][for more detail]{Toba_17b}.
We found that WISE1029 is located at a sequence of typical ULIRGs at similar redshifts \citep{Kilerci} on SFR-$M_*$ plane.
There is no evidence of quenched SF associated with the AGN activities.

\subsection{$L_{\rm FIR}$ and $L^\prime_{\rm CO}$ relation}
\label{FIR_CO}
Following that, we investigate the relation between FIR luminosity and CO luminosity of WISE1029.
FIR luminosity, $L_{\rm FIR}$ (40-120 $\micron$) is estimated using a SED fitting as described in Section \ref{MS}.
CO luminosities in units of K km s$^{-1}$ pc$^2$ are derived using the following relation provided by \cite{Solomon},
\begin{equation}
L^{\prime}_{\rm CO} = 3.25 \times 10^7 \,S_{\rm CO} \Delta v \,\nu_{\rm obs}^{-2}\,D_L^2\, (1+z)^{-3},
\end{equation}
where $S_{\rm CO} \Delta v$, $\nu_{\rm obs}$, and $D_L$ are the CO integrated flux density in units of Jy km s$^{-1}$, the observing frequency in GHz, and the luminosity distance in Mpc, respectively.
The resultant FIR, CO(2--1), and CO(4--3) luminosities are listed in Table \ref{Table}.
We found that the luminosity ratio of CO(4--3) and CO(2--1) lines ($r_{\rm 42}$) to be 0.61 $\pm$ 0.07.
When we use a ratio of peak flux instead of luminosity, $r_{\rm 42}$ = 0.57 $\pm$ 0.04.
We found that $r_{\rm 42}$ of WISE1029 is consistent with that of ULIRGs at $0.2 <z < 1.0$ \citep{Combes} within errors.

\begin{figure}
\centering
\includegraphics[width=0.45\textwidth]{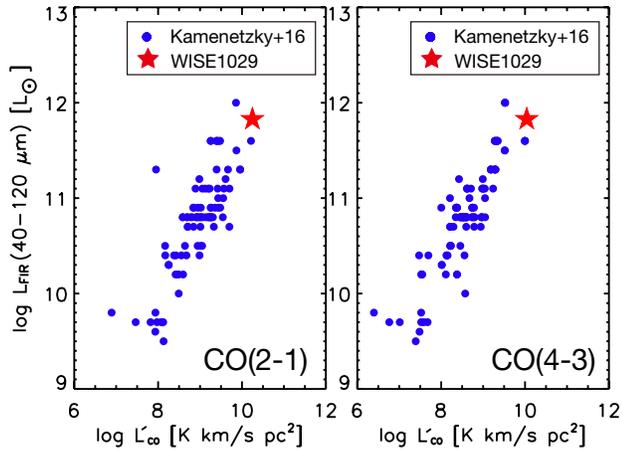}
\caption{The relation between FIR luminosity ($L_{\rm FIR}$ (40-120 $\micron$)) and CO luminosities for WISE1029 (red star) derived in this work and local (U)LIRGs (blue square) presented by \cite{Kamenetzky}.}
\label{LFIR_CO}
\end{figure}

It has been well-known that a good correlation exists between FIR and CO luminosities that is applicable for normal star forming galaxies and starburst galaxies \citep[e.g.,][]{Young,Greve,Michiyama}.
Figure \ref{LFIR_CO} shows $L_{\rm FIR}$ (40-120 $\micron$) and $L^{\prime}_{\rm CO}$ relation of WISE1029.
The relation for local IR galaxies at $z < 0.1$ reported by \cite{Kamenetzky} is also plotted.
We found that WISE1029 follows the correlation between FIR and CO luminosities of local ULIRGs, suggesting that the star formation efficiencies of molecular gas in WISE1029 is not strongly impacted by the presence of the AGN or the ionized outflow.

\subsection{Gas-to-dust mass ratio of WISE1029}
\label{Dust}
Finally, we investigate the dust and gas mass and gas-to-dust mass ratio of WISE1029.
Molecular (H$_2$) gas mass ($M_{\rm gas}$) is derived using the following formula:
\begin{equation}
M_{\rm gas} = \frac{\alpha_{CO}}{r_{\rm 21}}L^{\prime}_{\rm CO(2-1)},
\end{equation}
where $\alpha_{\rm CO}$ is the CO(1--0) to H$_2$ conversion factor while $r_{\rm 21}$ is ratio of $L^{\prime}_{\rm CO (2-1)}$ and $L^{\prime}_{\rm CO (1-0)}$.
In this work, we adopted $\alpha_{CO} = 0.8$ $M_{\sun}$ (K km $s^{-1}$)$^{-1}$ \citep{Downes} and $r_{\rm 21}$ = 0.91 \citep{Papadopoulos} that are typically used for (U)LIRGs.

In Section \ref{CO}, we mentioned that the broad component of CO lines are undetected.
Here we estimate the upper limit of the broad component of CO(2--1) luminosity and associated molecular mass.
Assuming that (i) outflowing gas associated with broad component is unresolved and (ii) FWHM of the broad component is 1000 km s$^{-1}$ that is the typical value for broad CO line of objects with strong molecular gas outflow \citep{Cicone_14}, we estimated the 3$\sigma$ upper limit of broad CO (2--1) line flux based on the RMS noise of the spectrum.
We obtained the broad CO (2--1) line flux is $<$ 0.02 Jy km s$^{-1}$, and associated luminosity is $\log\,L^{\prime}_{\rm CO(2-1)} < 8.1 $ K km s$^{-1}$ pc$^2$.
This means that even if a molecular outflow exists, its mass is $\log \, (M_{\rm gas}/M_{\sun}) <  8.1$ in order to remain undetected in our data.
Since this tight upper limit is$\sim$1/100 of the detected CO gas mass, with the current data is quite hard to detect the broad component if it exists.

Dust mass ($M_{\rm dust}$) is derived from the following formula:
\begin{equation}
M_{\rm dust} = \frac{D_L^2}{1+z}\, \frac{S (\nu_{\rm obs})}{\kappa_{\rm rest} B (\nu_{\rm rest}, T_{\rm dust})},
\end{equation}
where $S (\nu_{\rm obs})$ is flux density at observed frequency ($\nu_{\rm obs}$), $D_L$ is the luminosity distance, $\kappa_{\rm rest}$ is the dust mass absorption coefficient at rest frequency ($\nu_{\rm rest}$), and $B (\nu_{\rm rest}, T_{\rm dust}$) is the Planck function at temperature $T_{\rm dust}$ and at $\nu_{\rm rest}$.
We estimated dust mass at 850 $\micron$ using a dust absorption coefficient of $\kappa$ (850 $\micron$) =
0.383 cm$^2$ g$^{-1}$ \citep{Draine} and a dust temperature of $T_{\rm dust}$ = 40 K that are typical values for local ULIRGs \citep{Clements} and {\it Spitzer}-selected DOGs \citep{Melbourne}.
$S (\nu_{\rm obs})$ was inferred from the best-fit SED (see Figure \ref{SED}).

The resultant gas and dust mass are listed in Table \ref{Table}.
Dust grains formed in the interstellar medium (ISM) deplete metals from the gas phase. If dust is destroyed (e.g., by supernova shocks), metals are returned to the gas phase \citep[e.g.,][]{Draine_79,Jones,Savage}.
Hence, the gas-to-dust mass ratio ($M_{\rm gas}$/$M_{\rm dust}$) traces the chemical condition of galaxies and, although its value remains largely uncertain, it is an important measure to understand galaxy (particularly chemical) evolution.
It is generally thought that $M_{\rm gas}$/$M_{\rm dust}$ is $\sim$ 200--300 for local ULIRGs \citep[e.g.,][]{Seaquist} and that depends on metallicity \citep[e.g.,][]{Remy}. 
The estimated  $M_{\rm gas}$/$M_{\rm dust}$ of WISE1029 is 53 that is smaller than typical ULIRGs.
Since a prominent feature of DOGs is that they have relatively steep (red) continuum in the optical to NIR region compared to typical ULIRGs \citep[e.g.,][]{Toba_17c}, which is probably caused by a large amount of dust, one interpretation of our result is that WISE1029 is relatively dust rich object.

\section{Discussion} 
\label{Disc}

We discuss the interpretation of the result we obtained in this work.
We investigated the CO molecular gas properties of WISE1029, which shows a powerful ionized-gas outflow.
The derived IR luminosity of this DOG exceeds $10^{12}$ $L_{\sun}$ that is classified as an ULIRG.
The morphology of ULIRGs is often irregular probably due to the merger process \citep[e.g.,][]{Sanders,Murphy,Duc} and molecular gas in these ULIRGs are often disturbed \citep[e.g.,][]{Xu}. 
However, for WISE1029, we cannot see a significant sign of outflowing molecular gas in its CO spectra as well as moment 1 map, as described in Section \ref{RD}.

An interpretation of these results is that the ionized-gas outflow does not significantly affect the kinematics of molecular gas.
The above situation could occur if the ionized-gas is outflowing in a different direction from distribution of molecular gas, for example, along the path of least resistance perpendicular to the disk plane \citep[e.g.,][]{Harrison,McElroy}.
Since the moment 1 maps of our object is not inconsistent with molecular gas associated with the disk, the above scenario could be true in this DOG.
In this case, strong radiation from AGNs selectively affects the ionized-gas in the narrow line region (NLR) while the radiation pressure of the ionized-gas outflow does not significantly affect the ISM in the host galaxy at least for this DOG.
Further high resolution observations of ionized-gas  outflow and molecular gas are required to address the relative orientation of the ionized-gas outflow and molecular disk, and to determine whether the direction of the ionized-gas outflow is a key factor to understand how AGN-driven feedback affects the molecular gas properties.

It is common to see molecular outflows, if they are present, with velocities much slower than the ionized outflows \citep[e.g.,][]{Sun}, suggesting that AGN feedback could preferentially impact the diffused ionized gas more than the much denser molecular counterparts. 
On the other hand, many authors reported that many ULIRGs with ionized-gas outflows also show powerful molecular gas outflows \citep[e.g.,][and references therein]{Rupke}.
The case of WISE1029 is quite rare.
The only other example might be I Zw 1 (PG 0050+124), which shows high velocity ionized outflows \citep{Costantini}. 
This object does not show evidence of molecular gas outflow \citep{Cicone_14} although over 20\% of more extended emission might be missed by their observation.
The presence of a faint and broad CO component cannot be ruled with their data \citep[see][]{Cicone_14}.
Therefore, WISE1029 is likely to be an example of object in which there is no significant AGN feedback phenomenon despite of showing AGN-driven ionized-gas outflow.
This result shows that objects with powerful ionized-gas outflow do not always show powerful molecular gas outflow.

\acknowledgments 
The authors gratefully acknowledge the anonymous referee for a careful reading of the manuscript and very helpful comments.
This paper makes use of the following ALMA data: ADS/JAO.ALMA\#2015.1.00199.S. ALMA is a partnership of ESO (representing its member states), NSF (USA) and NINS (Japan), together with NRC (Canada), MOST and ASIAA (Taiwan), and KASI (Republic of Korea), in cooperation with the Republic of Chile. The Joint ALMA Observatory is operated by ESO, AUI/NRAO and NAOJ. 
Funding for SDSS-III has been provided by the Alfred P. Sloan Foundation, the Participating Institutions, the National Science Foundation, and the U.S. Department of Energy Office of Science. The SDSS-III web site is http://www.sdss3.org/.
SDSS-III is managed by the Astrophysical Research Consortium for the Participating Institutions of the SDSS-III Collaboration including the University of Arizona, the Brazilian Participation Group, Brookhaven National Laboratory, Carnegie Mellon University, University of Florida, the French Participation Group, the German Participation Group, Harvard University, the Instituto de Astrofisica de Canarias, the Michigan State/Notre Dame/JINA Participation Group, Johns Hopkins University, Lawrence Berkeley National Laboratory, Max Planck Institute for Astrophysics, Max Planck Institute for Extraterrestrial Physics, New Mexico State University, New York University, Ohio State University, Pennsylvania State University, University of Portsmouth, Princeton University, the Spanish Participation Group, University of Tokyo, University of Utah, Vanderbilt University, University of Virginia, University of Washington, and Yale University.
This publication makes use of data products from the Wide-field Infrared Survey Explorer, which is a joint project of the University of California, Los Angeles, and the Jet Propulsion Laboratory/California Institute of Technology, funded by the National Aeronautics and Space Administration. 
This research is based on observations with AKARI, a JAXA project with the participation of ESA.
Y.Toba and W.H.Wang acknowledge the support from the Ministry of Science and Technology of Taiwan (MOST 105-2112-M-001-029-MY3).
T.Nagao is financially supported by the Japan Society for the Promotion of Science (JSPS) KAKENHI (16H01101 and 16H03958). 
S.Komugi is supported by JSPS KAKENHI (15H02074).

\end{document}